\documentclass[12pt]{utarticle}
\usepackage{amsmath,amsfonts,pgfplots,bbm,mathrsfs}
\usepackage{epsfig,color,colortbl,tikz,pgfplots,hyperref}
\usepackage{youngtab}
\title{\vskip-0.5cm Light-Front Holographic QCD from\\ a Coherent State in String Theory}
\author{Harun Omer}
\oneaddress{homer@umassd.edu}
\Abstract{
\noindent 
It is shown that Light-Front Holographic Quantum Chromodynamics (LFHQCD) can be embedded in type II string theory based on the coherent state $e^{-\frac{1}{\lambda} L_{-1}}|h\rangle$. Salient features of hadron spectroscopy as known from LFHQCD carry over to string theory, notably the linear Regge trajectories of mesons and baryons in the chiral limit where $\lambda$ defines the QCD scale. From the LFHQCD perspective, this is an avenue to implement hadron spin. From the string perspective, string theory remains as a TOE with excited states generally at the Planck scale. Yet due to the unique properties of the coherent state, the mass scale of the tower of massive states on it is broken down from the Planck scale to the QCD scale, thereby reproducing properties of hadronic spectra. Furthermore, connection exists to gravitationally dressed excited states in AdS${}_3$.}
\begin{document}

\maketitle 
\section{Introduction}
String theory initially sparked enthusiasm for its ability to reproduce features of hadron spectroscopy  out of a rather simple action which does not appear specifically tuned towards exhibiting these features.
Following Veneziano's pioneering work in 1968 the dual resonance models were developed as a precursor of string theory which described $N$-point scattering amplitudes for mesons such as
$A(s,t)=\frac{\Gamma(-\alpha(s))\Gamma(-\alpha(t))}{\Gamma(-\alpha(s)-\alpha(t))}.$ Moreover, the string picture naturally explained the duality of the amplitude under the exchange of the $s$- and $t$-channel, $A(s,t)=A(t,s).$ In agreement with experimental results,  double-counting of dual amplitudes does not happen.
Mesons were understood rather intuitively as open strings with a quark and anti-quark attached to its two endpoints. Rotating strings possess angular momentum and vibrating strings are excited states, giving rise to a Regge trajectory of the form $J=\alpha_0+\alpha'M^2$. For the early development of string theory see for example~\cite{book_Rickles,book_dual_resonance}.
Yet at present, these striking properties have largely been given up on. With the development of Quantum Chromodynamics (QCD) following the work of Abdus Salam and Steven Weinberg, some features of the dual resonance models merged into  QCD, notably work on scattering amplitudes but also  the  string picture of flux tubes between quarks. Other features were largely abandoned, owing to unresolved problems.  String theory survived mainly for its other intriguing property, as a theory of quantum gravity, and no longer a theory for hadrons. The loss is significant. The scale of string theory became to be understood to be the Planck scale which reduced the interest in the string states solely to the massless spectrum, whose states are thought to acquire masses through small quantum corrections. Today's string theory models can not reproduce the Regge trajectory nor hadronic scattering amplitudes since both are associated with its massive states. In fact, it seems elusive to even in principle compute them for a given string vacuum, at least in the current state of string theory. Consequently, model-building has focused merely on reproducing generic features of the standard model, such as its gauge groups and matter content. In intersecting brane models, typically a stack of 'color' branes intersects a stack of 'weak' branes giving rise to an open string at the intersection which is interpreted as a quark. The perspicuous albeit admittedly oversimplified visualization of a meson as a spinning, vibrating string is lost and with it the simple understanding of confinement.

At present, a number of obstacles that caused interest to fade have already been overcome or appear less problematic. Compactified extra-dimensions are now omnipresent in string theory and within the field have come to be viewed as a necessary ingredient of a realistic theory.
The unphysical tachyonic ground state and the question of overall consistency of the theory has been resolved since supersymmetry has been incorporated and string theory has been proven to be anomaly free. More serious are the results of deep inelastic scattering experiments, which suggest that quarks have a point-like structure. The approach presented here connects to the  AdS/QCD\ duality which implies a semi-classical approximation with negligible quark masses and as such can not be expected to be able to model deep inelastic scattering. A further problem has been the unphysical unit intercept of the Regge trajectories in string theory. The lowest energy hadronic state in the chiral limit should be the massless spin 0 pion. This can be addressed by the compactification structure of the theory. Another fundamental
 question is how a mass scale other than the Planck scale enters the theory. This is addressed in this work. The approach presented embeds Light-Front Holographic QCD\ (LFHQCD) into string theory. Two review papers are~\cite{review1,review2}. The underlying quantum group is the 0+1 dimensional $OSp(2,1)$ group which essentially describes the hadronic spectrum via a confining harmonic oscillator potential. The phenomenological results carry over and are not addressed here in detail. As the name already indicates, light-front holographic QCD connects the underlying conformal theory to a dual on $AdS_2$ space. The holographic duality is not lost in this paper's construction. The coherent state in question has been associated with the trajectory of a particle along a geodesic in AdS space~\cite{AdS_coherent}. The analysis in this paper is entirely group theoretical. Familiarity with string theory at the level of standard textbooks such as~\cite{gsw} is assumed. 
\section{The Quantum Group of Light-Front Holographic QCD}
This section is essentially review. The quantum group of the type 2 superstring is the $\mathcal{N}=2$ superconformal algebra which is presumed to be familiar to the reader. We will be working in the NS sector. Its global subgroup with $\mathcal{N}=2$ supersymmetry corresponds to the group $OSp(1,2)$ on which superconformal quantum mechanics as well as LFHQCD is based. The bosonic subgroup of $OSp(1,2)$ is the conformal group in in 0+1 dimensions Conf$(\mathbb{R}^{0,1})$ which is isomorphic to $SL(2,\mathbb{R})$.
The algebra of the conformal group in 0+1 dimensions Conf$(\mathbb{R}^{0,1})$ is,
\begin{align}
[H,D]=iH \qquad [K,D]=-iK\qquad [H,K]=2iD.
\end{align}
At the level of the algebra the isomorphism to $SL(2,\mathbb{R})$ is realized by the identification,
\begin{align}
L_{-1}=H \qquad L_0=iD\qquad L_1=-K.\label{eq:isom1}
\end{align}
In solving the quantum model based on this algebra, de Alfaro, Fubini and Furlan~\cite{dAFF} noted that $H$ possesses a continuous spectrum and non-renormalizable eigenstates. Therefore it would be more practicable to choose a different operator to study the time evolution of the quantum system. They argued that any linear combination of the three generators $H_0=uH+vK+wD$ can be used as Hamiltonian so long as the resulting operator is compact. That is the case whenever $uw-v^2>0$.   The operator $H_0$ then generates the evolution of the quantum system $H_0|\psi(\tau)\rangle = i\frac{\partial}{\partial \tau}|\psi(\tau)\rangle$ with respect to a new parameter $\tau$ defined by,
\begin{align}
d\tau = \frac{dt}{u + v t + w t^2}.
\end{align}
.When identifying the generators with physical operators, one finds that the three generators all have different dimensions. 
Consequently the coefficients $u,v,w$ are dimensionful, despite the group being conformal. In that way, a scale enters the theory via the dimensionful linear coefficients. This paper presents an equivalent alternative perspective to this argument. If we consider a physical realization of the algebra in which the exponentiated generator is dimensionful, the scale  now derives directly from the coherent state $e^{-\frac{1}{\lambda}L_{-1}}|0\rangle$ with $\lambda$ ensuring the argument of the exponential is dimensionless. In any case, even in a unitary theory where $L_{-1}^{\dagger}=L_1$ the parameter $\lambda$ will be the scaling parameter of the theory as will become clear further below. Let us now proceed and select a Hamiltonian.
Following de Alfaro, Fubini and\ Furlan we choose,
\begin{align}
H_0=\frac{1}{2}\left(L_{-1}-\lambda^2 L_1\right).\label{eq:hamF}
\end{align}
To solve the system one defines the operators,
\begin{align}
H_{\pm 1}=\frac{1}{2}\left(L_{-1}+\lambda^2 L_1\pm 2\lambda L_{0}\right).\label{eq:ladderF}
\end{align}
which act as ladder operators,
\begin{align}
[H_0,H_{\pm1}]=\mp\lambda H_{\pm 1}\qquad [H_1,H_{-1}]=2\lambda H_0.\label{eq:scal_alg}
\end{align}
For $\lambda=1$ the usual $SL(2,\mathbb{R})$ algebra is restored. In the limit $\lambda \rightarrow 0$ the Hamiltonian $H_0$ reduces to $\frac{1}{2}L_{-1}$ but so do both of the ladder operators $H_1$ and $H_{-1}$. 
A sign inversion of $\lambda$ flips the roles of $H_{1}$ and $H_{-1}$. In the following it shall be assumed that $\lambda>0$. One can now find the associated tower of states using the lowest weight method. The lowest weight state $|\phi_h\rangle$ is annihilated by the destruction operator $H_1|\phi_h\rangle = 0$ and it is eigenstate to the operator acting as Hamiltonian $H_0$. For later convenience we write its eigenvalue as $\lambda h$, that is $H_0|\phi_h\rangle = \lambda h |\phi_h\rangle$. The application of a ladder operator on a state increases respective lowers the eigenvalue by one unit of $\lambda$, for example,
\begin{align}
H_0H_{-1}|\phi_h\rangle = ([H_{0},H_{-1}]+H_{-1}H_{0})|\phi_h\rangle=(\lambda H_{-1}+H_{-1}\lambda h)|\phi_h\rangle= (h+1)\lambda H_{-1}|\phi_h\rangle.
\end{align}
The bosonic sector of light-front holographic QCD is based on this spectrum. The procedure can be applied to the supersymmetric extension to
$OSp(1,2)$~\cite{fubinirabinovici}.
\section{Realization in Terms of the  Coherent State}
This section addresses how the above quantum group can be embedded into string theory. Specifically, it needs to be established how the state  $|\phi_h\rangle$ relates to the string ground state $|h\rangle$. The string ground is annihilated by $L_1$ whereas $|\phi_h\rangle$ is annihilated by $H_1$:
\begin{align}
L_1|h\rangle=0 \qquad H_1|\phi_h\rangle=0
\end{align}
Both conditions are equivalent if we can relate,
\begin{align}
VL_1V^{-1}\propto H_1 \qquad |\phi_h\rangle=V|h\rangle.
\end{align}
With the Baker-Hausdorff lemma it is easy to show that $V=e^{-\frac{1}{\lambda} L_{-1}}$ achieves that objective:
\begin{align}
 VL_1V^{-1}&=\textstyle\frac{2}{\lambda^2} H_1.
\end{align}
Making use of $L_0|h\rangle =h |h\rangle$ one can verify that the state $|\phi_h\rangle$ constructed in this way is indeed eigenstate of $H_0$ and has eigenvalue $\lambda h$,
\begin{align}
H_0|\phi_h\rangle=VV^{-1}H_{0}V|h\rangle=V(\lambda L_0-\frac{1}{2}\lambda^2L_1)|h\rangle=V\lambda h|h\rangle=\lambda h|\phi_h\rangle.\label{eq:hamil}  
\end{align}
These two conditions together with the operator algebra are sufficient to show that the spectra are identical. The above one-line expression is central to this paper and it is worth looking at it in detail. It is not possible to arbitrarily modify $V$ to obtain similar constructions. Generically, $V^{-1}H_0V$ will result in some linear combination of $L_{-1}$, $L_0$ and $L_1$. If $L_{-1}$ appeared in it, the eigenvalue spectra of $H_0$ and $L_0$ could not be related. If solely $L_0$ appeared, the connection between $H_0$\ and $L_0$ would be trivial. In Eq.~(\ref{eq:hamil}) the specific form of $V$ causes the $L_{-1}$-term to vanish, causes the $L_0$-term to pick up the scaling factor $\lambda$ and causes the $L_1$-term to appear as the only term which drops out by virtue of annihilating $|h\rangle$.  

Let us now turn to supersymmetry. The supercharges,
\begin{align}
\begin{array}{rcl}
R^+&\equiv& G^-_{-\frac{1}{2}}+\lambda G^-_{\frac{1}{2}}\\
R^-&\equiv& G^+_{-\frac{1}{2}}-\lambda G^+_{\frac{1}{2}}\\
\end{array}
\end{align}
satisfy the supersymmetry conditions,
\begin{align}
\{R^+,R^-\}=2H_0=L_{-1}-\lambda^{2}L_{1}-2\lambda J_{0}\qquad \{R^{-},R^{-}\}=0 \qquad \{R^{+},R^{+}\}=0.\label{eq:susyR}
\end{align}
In contrast to the purely bosonic discussion the Hamiltonian picks up an additional $J_0$ term from the supercurrent. Recall that $J_0$ commutes with all bosonic operators.
In the supersymmetric theory, each eigenstate $|\psi\rangle$ of $H_0$ has the supersymmetric fermionic partner state $R^{+}H_0|\psi\rangle$ with the same eigenvalue. At the ground state level, however, supersymmetry is broken. This can be seen by acting on the ground state with the supersymmetry operator,
\begin{align}
\begin{array}{rcl}
R^{+}|\phi_h\rangle&=&e^{-\frac{1}{\lambda}L_{-1}}e^{\frac{1}{\lambda}L_{-1}}(G^-_{-\frac{1}{2}}+\lambda G^-_{\frac{1}{2}})e^{-\frac{1}{\lambda}L_{-1}}|h\rangle\\
&=&e^{-\frac{1}{\lambda}L_{-1}}(G^-_{-\frac{1}{2}}+\lambda( G^-_{\frac{1}{2}}-\frac{1}{\lambda}G^-_{-\frac{1}{2}}))|h\rangle\\
&=&0.
\end{array}
\end{align}
Since $G^{-}_{-\frac{1}{2}}$ precisely cancels out in the second line, $R^{+}$ annihilates the lowest state. Generally it is understood that in spontaneous symmetry breaking degenerate ground states exist of which one has to be chosen. The symmetry breaking here is realized by
the bosonic state $|\phi_h\rangle$ not having a non-trivial supersymmetric partner. Of course, if the lowest state is projected out as in string theory when applying the GSO projection, this discussion does not apply.

A few remarks on unitarity are in order. The attentive reader will have noticed that $H_{m}\ne H_{-m}^{\dagger}$ and $R^{+\dagger}\ne R^{-}$. Typically a Hamiltonian is taken to be hermitian, which is a sufficient condition to ensure it has real eigenvalues and real expectation values.
But not ever Hamiltonian needs to be hermitian. Perhaps most prominently, even the Klein-Gordon equation does not have a hermitian Hamiltonian. An operator that is related to a hermitian operator by a similarity transformation will have the same eigenvalues as the hermitian operator. In our case the relationship between the Hamiltonian $H_0$ and the self-adjoint $L_0$ as given in Eq.~(\ref{eq:hamil}) ensures the reality of the eigenvalues. References~\cite{NONHERMIT1} and~\cite{NONHERMIT2} discuss non-nermitian Hamiltonians from a physical respectively mathematical perspective. The specific Lie algebra in question is discussed in~\cite{NONHERMIT3}. Irrespective of such formal considerations, the ladder structure in Eq.~(\ref{eq:scal_alg}) explicitly shows that all eigenvalues are real provided that $\lambda$ is real. That being said, an analogous construction with Hermitian operators satisfying $H_0^{\dagger}=H_0$ and $H_{-1}^{\dagger}=H_1$ is possible.\footnote{Then,
\begin{align}
\begin{array}{ll}
H_0 &\rightarrow -\frac{1}{1-\lambda^2}((1+\lambda^2)L_{0}+\lambda(L_{-1} +L_1)),\\
H_{-1} &\rightarrow-\frac{1}{1-\lambda^2}(2\lambda L_{0}+L_{-1} +\lambda^2L_1),\\
H_{+1} &\rightarrow-\frac{1}{1-\lambda^2}(2\lambda L_{0}+\lambda^2L_{-1} +L_1).
\end{array}
\end{align}
These operators are left invariant up to an overall sign inversion when $\lambda\rightarrow \frac{1}{\lambda}$.\ Note the similarity to T-duality. But regardless of any physical interpretation of these operators, here $\lambda$ must be dimensionless.
}
\section{Oscillator Representation and Hadron Angular\ Momentum}
The 0+1 dimensional superconformal algebra to which the LFHQCD\ model effectively reduces, does not intrinsically incorporate spin.  This is not surprising since spin relates to rotations in space and requires more than a single dimension.   For integer spin, the spin interaction has been obtained by embedding the light-front Hamiltonian in AdS, which results in an expression for the dilaton profile and thus the spin term. For half-integer spin the dilaton profile in the action can be rotated away, giving no additional constraints. In the AdS treatment of Rarita-Schwinger equations, a spin operator for half-integer spin is also absent~\cite{LFHQCD_3,LFHQCD_6}.
Supersymmetry has been used to deduce the spin interactions for baryons~\cite{LFHQCD4}.  Ultimately, a $\lambda \frac{S}{2}$ spin term was added by hand to the $OSp(1,2)$ Hamiltonian. We shall now aim to introduce angular momentum in our model. The key to that end is the oscillator representation. The $SL(2,\mathbb{R})$ generators can be written in an oscillator basis as,
\begin{align}
L_0=\frac{1}{4}\left( a a^{\dagger}+a^{\dagger}a\right)\qquad L_{1}=\frac{1}{2}a a\qquad L_{-1}=\frac{1}{2}a^{\dagger}a^{\dagger},\label{eq:oscia}
\end{align}
subject to $[a,a^{\dagger}]=1$. The commutators of the symmetry generators with $a$ and $a^{\dagger}$ are,
\begin{align}
[L_0,a]=-\frac{1}{2}a\qquad [L_0,a^{\dagger}]=\frac{1}{2}a^{\dagger}\qquad[L_1,a^{\dagger}]=a\qquad [L_{-1},a]=-a^{\dagger}.
\end{align}
Since the $H_m$ obey the same symmetry as the $L_m$ apart from the scaling factor, they must possess an equivalent oscillator representation. Indeed, when expressed in terms of $a$ and $a^{\dagger}$, the $H_m$ all factor,
\begin{align}
H_{0}&=\frac{1}{4}\left(A_{-} A_{+}+A_{+}A_{-}\right)\qquad H_{1}=\frac{1}{2} A_{-}A_{-}\qquad H_{-1}=\frac{1}{2} A_{+}A_{+},\label{eq:opbigA}
\end{align}
where,
\begin{align}
A_{\pm}=\frac{1}{\sqrt{2}}\left(-a^{\dagger}\pm\lambda a\right).\label{eq:oscbigA}
\end{align}
While the operators take the same form, this is not an automorphism on the oscillator algebra. The difference is that the newly introduced oscillator modes satisfy $[A_{-},A_{+}]=\lambda$. 

One can now proceed by attaching a space-time label to each mode so that  $[A^{\mu}_{-},A^{\nu}_{+}]=\lambda g^{\mu\nu}$. The Hamiltonian becomes $H_0\rightarrow \frac{1}{4}g_{\mu\nu}\left(A^{\mu}_{+}A^{\nu}_{-}+A^{\nu}_{+}A^{\mu}_{-}\right)$ and the ladder operators are modified analogously. It is tempting to construct the state $A^{\mu}_{+}|\phi_h\rangle$ which transforms as a vector and identify it with a spin 1 state. Indeed this state is an eigenstate to $H_0$ and raises the eigenvalue by half a unit of $\lambda$,
\begin{align}
\textstyle H_{0}A^{\mu}_{+}|\phi_h\rangle=(A^{\mu}_{+}H_0+ \frac{1}{2}\lambda A^{\mu}_{+})|\phi_h\rangle=\lambda(h+\frac{1}{2})A^{\mu}_{+}|\phi_h\rangle.
\end{align}
This half-unit increase precisely matches the spin term $\lambda\frac{S}{2}$ from the LFHQCD models. Higher spin states can be constructed by acting on the state with additional operators. This certainly looks promising, yet spin is more than a space-time label. We will in the following work with  the oscillator representation of the $\mathcal{N}=2$ superstring where the notion of spin is well-defined.

Recall that the standard oscillator representation of the bosonic string is,
\begin{align}
L_m&=\frac{1}{2}\sum_{n=-\infty}^{\infty}:\alpha^{\mu}_{m-n}\alpha_{n,\mu}:+h\delta_{m,0}. \end{align}
The normal ordering constant $h$ only appears in $L_0$ and is known to be $-1$ for the open string in the NS-sector. The modes satisfy,
\begin{eqnarray}
[\alpha^{\mu}_n,\alpha^{\nu}_m]=n \eta^{\mu\nu}\delta_{n+m,0}\qquad \alpha^{\mu\dagger}_n=\alpha^{\mu}_{-n},
\end{eqnarray}
where $\alpha^{\mu}_{n}|h\rangle=0$ for $n>0$ and $\alpha_0^{\mu}$ commutes with all operators.
Let us express the first excited state in terms of oscillators,
\begin{align}
\begin{array}{rcl}
H_{-1}|\phi_h\rangle&=& VV^{-1}H_{-1}V|h\rangle\\
&=&V\left(2L_{-1}+\frac{1}{2}\lambda^2 L_1-2 \lambda L_{0}\right)|h\rangle\\
&=&V(2L_{-1}-2\lambda h)|h\rangle\\
&=&2V(\alpha^{\mu}_0\alpha_{-1,\mu}-\lambda h)|h\rangle
\end{array}
\end{align}
It is straightforward to confirm that this state is indeed eigenstate to $H_0$ with eigenvalue $\lambda(h+1)$. In the above state, all space-time indices are contracted.  
An analogous state with one free space-time index would be $V(\alpha^{\mu}_{-1}-\frac{1}{2}\lambda \alpha_0^{\mu})|h\rangle$. An explicit calculation confirms that the state is indeed an eigenstate with the same eigenvalue:
\begin{align}
\begin{array}{rcl}
H_0 V(\alpha^{\mu}_{-1}-\frac{1}{2}\lambda \alpha^{\mu}_0)|h\rangle&=&VV^{-1}H_0V(\alpha^{\mu}_{-1}-\frac{1}{2}\lambda \alpha^{\mu}_0)|h\rangle\\
&=&V(\lambda L_0-\frac{1}{2}\lambda^2L_1)(\alpha^{\mu}_{-1}-\frac{1}{2}\lambda \alpha^{\mu}_0)|h\rangle\\
&=&\lambda(1+h) V(\alpha^{\mu}_{-1}-\frac{1}{2}\lambda\alpha^{\mu}_0)|h\rangle.
\end{array}\label{eq:evspin}
\end{align}
The simplest bosonic state with two free space-time indices is $V(\alpha^{\mu}_{-1}-\frac{1}{2}\lambda \alpha^{\mu}_0)(\alpha^{\nu}_{-1}-\frac{1}{2}\lambda \alpha^{\nu}_0)|h\rangle$ and has eigenvalue $\lambda(2+h)$. States with more free space-time indices are constructed analogously. For a more formal analysis of spin, one has to look at the eigenstates of the spin operator whose bosonic part is given by,
\begin{align}
S_{\text{bos}}^{\mu\nu}=\mathrm{-i}\sum_{k=1}^{\infty}\frac{1}{k}(\alpha^{\mu}_{-k}\alpha^{\nu}_k-\alpha^{\nu}_{-k}\alpha^{\mu}_{k})
\end{align}
Since the spin operator commutes with $L_m$ and with $V=e^{-\frac{1}{\lambda}L_{-1}}$, one can directly draw from the results found in any standard text on string theory. For instance it is known that the level 2 state $(s_{ij}\alpha^{i}_{-1}\alpha^{j}_{-1}+v_{i}\alpha^{i}_{-2})|h\rangle$
 combines into a spin 2 representation where $s_{ij}$ is a symmetric matrix and $v_i$ a vector. In order to construct
the analogues of the higher level modes $\alpha_{-k}$ with $k>1$ as in Eq.~(\ref{eq:evspin}), the following commutator is useful:
\begin{align}
[\lambda L_{0}-\frac{1}{2}\lambda^2 L_1,\sum_{k=0}^n\left(-\frac{\lambda}{2}\right)^{k}\begin{pmatrix}n\\k\end{pmatrix}\alpha^{\mu}_{-n+k}]=\lambda n \sum_{k=0}^n\left(-\frac{\lambda}{2}\right)^{k}\begin{pmatrix}n\\k\end{pmatrix}\alpha^{\mu}_{-n+k}.
\end{align}
From this relation it is straightforward to obtain the level 2 state with one free space-time label as $V(\alpha^{\mu}_{-2}-\lambda\alpha^{\mu}_{-1}+\frac{1}{4}\lambda^2\alpha^{\mu}_{0})|h\rangle$ and the level 3 state
$V(\alpha^{\mu}_{-3}-\frac{3}{2}\lambda\alpha^{\mu}_{-2}+\frac{3\lambda^2}{4}\alpha^{\mu}_{-1}-\frac{\lambda^3}{8}\alpha^{\mu}_{0})|h\rangle$.

Let us now turn to the world-sheet fermions. In terms of the fermionic modes the Virasoro generators can be expressed as,
\begin{align}
\displaystyle L_m=\frac{1}{4}\sum_{r\in \mathbb{Z}+\frac{1}{2}}(2r-m):\psi^{\mu}_{-r}\psi_{m+r,\mu}:+h\delta_{m,0}.
\end{align}
The normal ordering constant for the worldsheet fermions is known to be $-\frac{1}{2}$. and the oscillators satisfy,
\begin{align}
\{\psi^{\mu}_r,\psi^{\nu}_s\}=\eta^{\mu\nu}\delta_{r+s,0}\qquad \psi^{\mu\dagger}_r=\psi^{\mu}_{-s}.
\end{align}
The level $\frac{1}{2}$, level $\frac{3}{2}$ and level $\frac{5}{2}$ creation operators with one free space-time index  are $V\psi^{\mu}_{-\frac{1}{2}}$, $V(\psi^{\mu}_{-\frac{3}{2}}-\frac{1}{2}\lambda \psi^{\mu}_{-\frac{1}{2}}$) and $V(\psi^{\mu}_{-\frac{5}{2}}-\lambda \psi^{\mu}_{-\frac{3}{2}}+\frac{1}{4}\lambda ^{2}\psi^{\mu}_{-\frac{1}{2}})$ respectively.
\section{Phenomenology}
\begin{figure}
\begin{center}
   \pgfplotsset{
    small,
    legend style={
        at={(0.01,0.01)},
        anchor=south west,
    },
   }%
   \begin{tikzpicture}[baseline]
   \begin{axis}[
   xmax=3,xmin=0,
   ymin= 0,ymax=3,
   xlabel=\emph{Angular Momentum $\qquad J$},ylabel=\emph{$\frac{\alpha'M^2}{\lambda}$ [GeV$^2$]},
   xtick={0,1,2,...,3},
   ytick={0,1,...,3},
   legend pos=south east,
   ]
   \addplot coordinates{(0,0) (1,1) (2,2) (3,3) (4,4)};
   \addplot coordinates{(0,1) (1,2) (2,3) (3,4)};
   \addplot coordinates{(0,2) (1,3) (2,4)};
   \addplot coordinates{(0,3) (1,4)};
   \addplot coordinates{(0,4)};
    \end{axis}
    \end{tikzpicture}%
    ~%
\end{center}
\caption{Regge trajectories}
\label{fig:regtraj}
\end{figure}
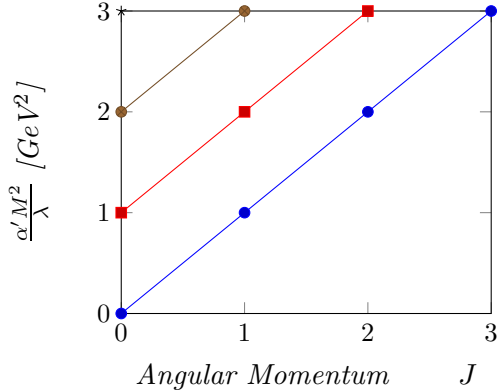

The GSO projection removes the states whose fermion number is zero, including the negative mass and zero mass bosonic states. The level $\frac{1}{2}$ state corresponding to $V\psi^{\mu}_{-\frac{1}{2}}$ is the massless state. It transforms as a vector, whereas for a hadronic theory the massless state should correspond to the spin 0 pion. Furthermore,  the mass scale $\lambda$ contained in $V=e^{-\frac{1}{\lambda}L_{-1}}$
is not universal but should differ for the hadron families. Both issues are resolved if the states live on D-branes. In fact, there must be D-branes present since all states that were considered have been open string states. From the perspective of a D-brane extended in the $x_4$-direction for example, $V\psi^{4}_{-\frac{1}{2}}$ behaves like a massless spin 0 object and can be identified with the pion. On another D-brane the scale $\lambda$ can be different. Together with the higher-level states which are unaffected by the GSO projection, one recovers the Regge trajectories shown in Fig.~\ref{fig:regtraj}. An open question remain the $\rho$-meson trajectories. While states do exist which match with a $\frac{S}{2}$ shift of the mass spectrum, these states have even world-sheet fermion number and are projected out when supersymmetry is unbroken.  


\begin{thebibliography}{99}
\bibitem{book_Rickles} D.~Rickles, {\it "A brief history of string theory: From dual models to M-theory"}, Springer, Berlin/Heidelberg (2014).
\bibitem{book_dual_resonance} P.~Frampton, {\it "Dual Resonance Models and Superstrings"}, \url{https://doi.org/10.1142/0249}, World Scientific, Singapore, (1986).
Rev. D \textbf{91}, 085016 (2015) {\tt [arXiv:1501.00959 [hep-th]]}.
\bibitem{review1} S.J.~Brodsky, {\it "Hadron Spectroscopy and Dynamics from Light-Front Holography and
Superconformal Algebra"},  Few-Body Syst \textbf{59}, 83 (2018) {\tt [arXiv:1802.08552 [hep-ph]]}.
\bibitem{review2} L.~Zhou, H.G.~Dosch, {\it "A very Practical Guide to Light Front Holographic QCD"}, {\tt [arXiv:1801.00607 [hep-ph]]}.
\bibitem{AdS_coherent} P.~Caputa, D.~Ge, {\it "Entanglement and geometry from subalgebras of the Virasoro algebra"},  J. High Energ. Phys. \textbf{2023}, 159 (2023). {\tt [arXiv:2211.03630 [hep-th]]}.
\bibitem{gsw} M.~Green, J.~Schwarz, E.~Witten, {\it "Superstring theory"}, Vol 1, Cambridge University Press (1988).
\bibitem{dAFF} V. de Alfaro, S. Fubini, G. Furlan, {\it "Conformal invariance in quantum mechanics"}, Nuovo Cim. A 34, 569 (1976).
\bibitem{fubinirabinovici} S.~Fubini, E.~Rabinovici, {\it "Superconformal quantum mechanics"}, Nucl. Phys. \textbf{B} 245, 17 (1984).
\bibitem{LFHQCD2}  H. G. Dosch, G. F. de T\'eramond, S. J. Brodsky, {\it "Superconformal baryon-meson symmetry and light-front holographic QCD"}, Phys. Rev. D \textbf{91}, 085016 (2015).
\bibitem{NONHERMIT1} C.M.~Bender, D.C.~Brody, H.F.~Jones, {\it "Must a Hamiltonian be Hermitian?"} Amer. J. Phys. 71 (2003), 1095-1102 {\tt [arXiv:0303005 [hep-th]]}.
\bibitem{NONHERMIT2} N.~Bebiano, J.~da~Providencia, J.P.~da~Providencia, {\it "Classes of non-hermitian operators with real eigenvalues"}. Electronic Journal of Linear Algebra 21 (2010) pp. 98-109.
\bibitem{NONHERMIT3} N.~Amaouche, M.~Sekhri, R.~Zerimeche, M~.Maamache,J.-Q.~Liang, {\it "Non-Hermitian Hamiltonian beyond PT symmetry for time-dependent $SU(1,1)$
and $SU(2)$ systems"}, Physics Open, Vol 13 (2022), 100126.
\bibitem{LFHQCD_3} G. F. de T\'eramond, S. J. Brodsky, {\it "Kinematical and Dynamical Aspects of Higher-Spin Bound-State Equations in Holographic QCD"}, Phys. Rev. \textbf{D} 87, 075005 (2013) {\tt [arXiv:1301.1651 [hep-ph]]}.
\bibitem{LFHQCD_6} S. J. Brodsky, G. F. de T\'eramond, H. G. Dosch, J. Erlich, {\it "Light-front holographic QCD and emerging confinement"}, Phys. Rep. 584, 1 (2015) {\tt [arXiv:1407.8131 [hep-ph]]}.
\bibitem{LFHQCD4} S. J. Brodsky, G. F. de T\'eramond, H. G. Dosch, C. Lorc\'e, {\it "Universal effective hadron dynamics
from superconformal algebra"}, Phys. Lett. B \textbf{759}, 171 (2016) {\tt [arXiv:1604.06746 [hep-ph]]}.
\end{thebibliography}
\end{document}